\documentclass[12pt]{article}
\usepackage{times}
\usepackage{geometry}
\usepackage[utf8]{inputenc}
\usepackage{enumitem,amssymb}
\usepackage{ragged2e}
\newlist{thematic}{itemize}{8}
\setlist[thematic]{label=$\square$}
\usepackage{pifont}
\newcommand{\cmark}{\ding{51}}%
\newcommand{\done}{\rlap{$\square$}{\raisebox{2pt}{\large\hspace{1pt}\cmark}}%
\hspace{-2.5pt}}

\usepackage{deluxetable} 
\usepackage{natbib} 
\usepackage{natbib} 
\usepackage{aas_macros} 
\usepackage{graphicx} 
\usepackage{wrapfig}

\begin{document}
\raggedright
\huge
Astro2020 Science White Paper \linebreak

Star Formation in Different Environments: The Initial Mass Function\linebreak
\normalsize

\noindent \textbf{Thematic Areas:} \hspace*{60pt} $\square$ Planetary Systems \hspace*{10pt} $\done$ Star and Planet Formation \hspace*{20pt}\linebreak
$\square$ Formation and Evolution of Compact Objects \hspace*{31pt} $\square$ Cosmology and Fundamental Physics \linebreak
  $\square$  Stars and Stellar Evolution \hspace*{1pt} $\done$ Resolved Stellar Populations and their Environments \hspace*{40pt} \linebreak
  $\square$    Galaxy Evolution   \hspace*{45pt} $\square$             Multi-Messenger Astronomy and Astrophysics \hspace*{65pt} \linebreak
  
\textbf{Principal Author:}

Name: Matthew Hosek Jr.
 \linebreak						
Institution: University of California at Los Angeles
 \linebreak
Email: mwhosek@astro.ucla.edu
 \linebreak
 
\textbf{Co-authors:} (names and institutions)
  \linebreak
Jessica R. Lu (University of California at Berkeley) \linebreak
Morten Andersen (Gemini Observatory) \linebreak
Tuan Do (University of California at Los Angeles) \linebreak
Dongwon Kim (University of California at Berkeley) \linebreak
Nicholas Z. Rui (University of California at Berkeley) \linebreak
Peter Boyle (University of California at Berkeley) \linebreak
Benjamin F. Williams (University of Washington) \linebreak
Sukanya Chakrabarti (Rochester Institute of Technology) \linebreak
Rachael L. Beaton (Princeton University) \linebreak
\newline
\textbf{Abstract:}
\newline
\newline
The stellar initial mass function (IMF) is a fundamental property of star formation, offering key insight into the physics driving the process as well as informing our understanding of stellar populations, their by-products, and their impact on the surrounding medium. While the IMF appears to be fairly uniform in the Milky Way disk, it is not yet known how the IMF might behave across a wide range of environments, such as those with extreme gas temperatures and densities, high pressures, and low metallicities. We discuss new opportunities for measuring the IMF in such environments in the coming decade with JWST, WFIRST, and thirty-meter class telescopes. For the first time, we will be able to measure the high-mass slope and peak of the IMF via direct star counts for massive star clusters across the Milky Way and Local Group, providing stringent constraints for star formation theory and laying the groundwork for understanding distant and unresolved stellar systems. 

\pagebreak
{\bf \large Introduction} 
\newline
Star formation has played a critical role in shaping the Universe we observe today.
A fundamental property of star formation is the Initial Mass Function (IMF), which 
describes the distribution of stellar masses created in a star-forming event. 
The IMF provides key insights into the underlying physics driving star formation \citep[e.g.][]{Krumholz:2014ne, Offner:2014vn}, 
and is a vital ingredient in many areas of astrophysics, 
such as star formation over cosmic time \citep[e.g.][]{Narayanan:2012tg, Ferre-Mateu:2013qo}, 
the mass assembly and evolution of galaxies \citep[e.g.][]{Clauwens:2016tw, Gutcke:2019jw}, 
compact object production and merger rates \citep[e.g.][]{2017MNRAS.467..524B, Mapelli:2018qf}, 
and stellar feedback \citep[e.g.][]{Dale:2015kl}.
 \newline 

Despite its importance, we still lack a model of star formation that 
predicts the IMF of a stellar population produced by a given molecular cloud. 
Observations suggest that the IMF is fairly uniform within the
Milky Way disk and local solar neighborhood \citep[][for review]{Bastian:2010dp}.
However, these populations span a limited range of environmental conditions. 
There is now evidence that the IMF varies in more extreme environments such as in the Galactic Center \citep[e.g.][]{Lu:2013wo}, 
the most massive elliptical galaxies \citep[e.g.][]{van-Dokkum:2010uk}, 
or the least luminous Milky Way satellites \citep[e.g.][]{Geha:2013ye}.
These claims of a non-standard IMF are debated and the underlying astronomical measurements cannot easily be improved upon with current space-based and ground-based observatories.
\newline

We discuss new opportunities for investigating the IMF across a wide range of 
environments in the coming decade. \emph{With advanced observing
facilities such as the James Webb Space Telescope (JWST), Wide-Field Infrared Survey Telescope (WFIRST), 
and 30m class telescopes, we can precisely measure the IMF down to 
$\leq$0.2~M$_{\odot}$ via star counts 
for star clusters across the Milky Way and Local Group.}
For the first time, we will be able to directly measure both the high-mass slope and peak mass of the
IMF for these populations and characterize how they vary with environment. 
This allows us to distinguish between different physical processes that 
influence star formation and inform how we should interpret
observations of unresolved stellar systems.
\newline

{\bf \large Current Status: IMF Variations with Environment?} 
\newline
The quest for a predictive theory of star formation as a function of initial conditions (metallicity, density, external pressure) is far from over.  
Most observations reveal a ``universal" IMF, and, not surprisingly, many diverse models are able to ``explain" it. 
Under intense scrutiny, most claims of detecting a non-universal IMF wither \citep[e.g.][]{Bastian:2010dp, Luhman:2018jl}, though there are some unexplained surprises.  
Some of the most persistent claims are from IMF measurements via indirect techniques (population synthesis, dynamics, and strong lensing) toward giant elliptical galaxies \citep[e.g.][]{van-Dokkum:2010uk, Treu:2010dt, Conroy:2012fk, Cappellari:2012zh, La-Barbera:2013wy, Spiniello:2014oq, Martin-Navarro:2015kq, Conroy:2017qc, van-Dokkum:2017bx, La-Barbera:2017kg, Parikh:2018uq}. 
These studies suggest that the IMF becomes increasingly bottom-heavy (i.e., an overabundance of low-mass stars) with increasing galaxy velocity dispersion and/or $\alpha$-element enhancement.
However, concerns have been raised about the impact of systematics on these analyses, such as elemental abundance gradients \citep[e.g.][]{McConnell:2016am, Zieleniewski:2017ye} or galaxy mass modeling \citep[e.g.][]{Leier:2016jo}. 
In addition, the internal consistency of IMF determinations using these indirect methods has yet to be established \citep[e.g.][]{Newman:2017ay}.
\newline

On the other hand, direct star counts have found top-heavy IMFs (i.e. an overabundance of high-mass stars) for two massive clusters at the Galactic Center \citep[][]{Lu:2013wo, Hosek:2019kk} and the 30 Dorodus starburst region in the LMC \citep{Schneider:2018ux}. 
These clusters are thought to form in significantly different conditions than typical clusters in the Milky Way disk; for example, the Galactic Center 
has been shown to exhibit similar gas densities, temperatures, and kinematics as starburst galaxies \citep{Kruijssen:2013lh, Ginsburg:2016nr}. 
On the other end of the spectrum, some ultra-faint dwarf galaxies have also been reported with top-heavy IMFs, perhaps due to the low metallicity or overall stellar mass of these systems \citep{Geha:2013ye, Gennaro:2018rf}, though others seem more consistent with the Milky Way disk \citep{Gennaro:2018wc}. 
Such direct measurements of the IMF face challenges as well, including sample contamination, stellar multiplicity, and limited stellar mass ranges. 
Unfortunately, nearby and better-studied star forming regions are poor analogues for the diverse star forming events at extreme temperatures, densities, pressures and low metallicities which may characterize environments in the early Universe, galactic nuclei, and merger events. 
\newline


{\bf \large The Next Step: Direct IMF Measurements Across Different Environments}
\begin{figure}
   \vspace{-1cm}
    \centering
    \begin{minipage}{0.48\textwidth}
        \includegraphics[scale=0.31]{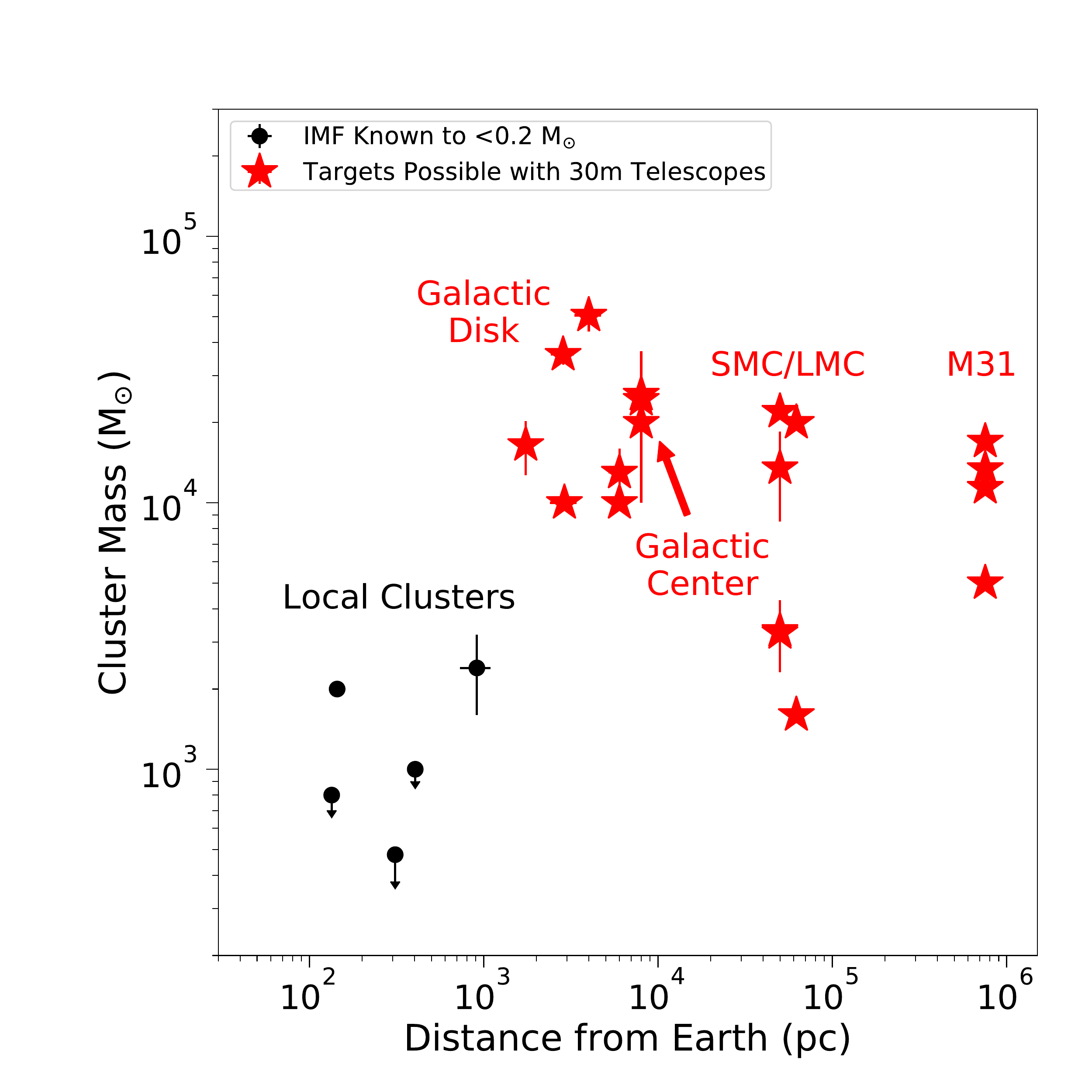}
        \caption{Examples of young and massive clusters for which we can measure the IMF to $\leq$ 0.2 M$_{\odot}$ with 30m class telescopes (red stars) compared to local star forming regions where this measurement has already been made (black points). The substantial gain in both spatial resolution and sensitivity allows us to probe a wide range of environments for the first time.}
        \label{fig:sample}
    \end{minipage}\hfill
    \begin{minipage}{0.48\textwidth}
        \centering
        \includegraphics[scale=0.31]{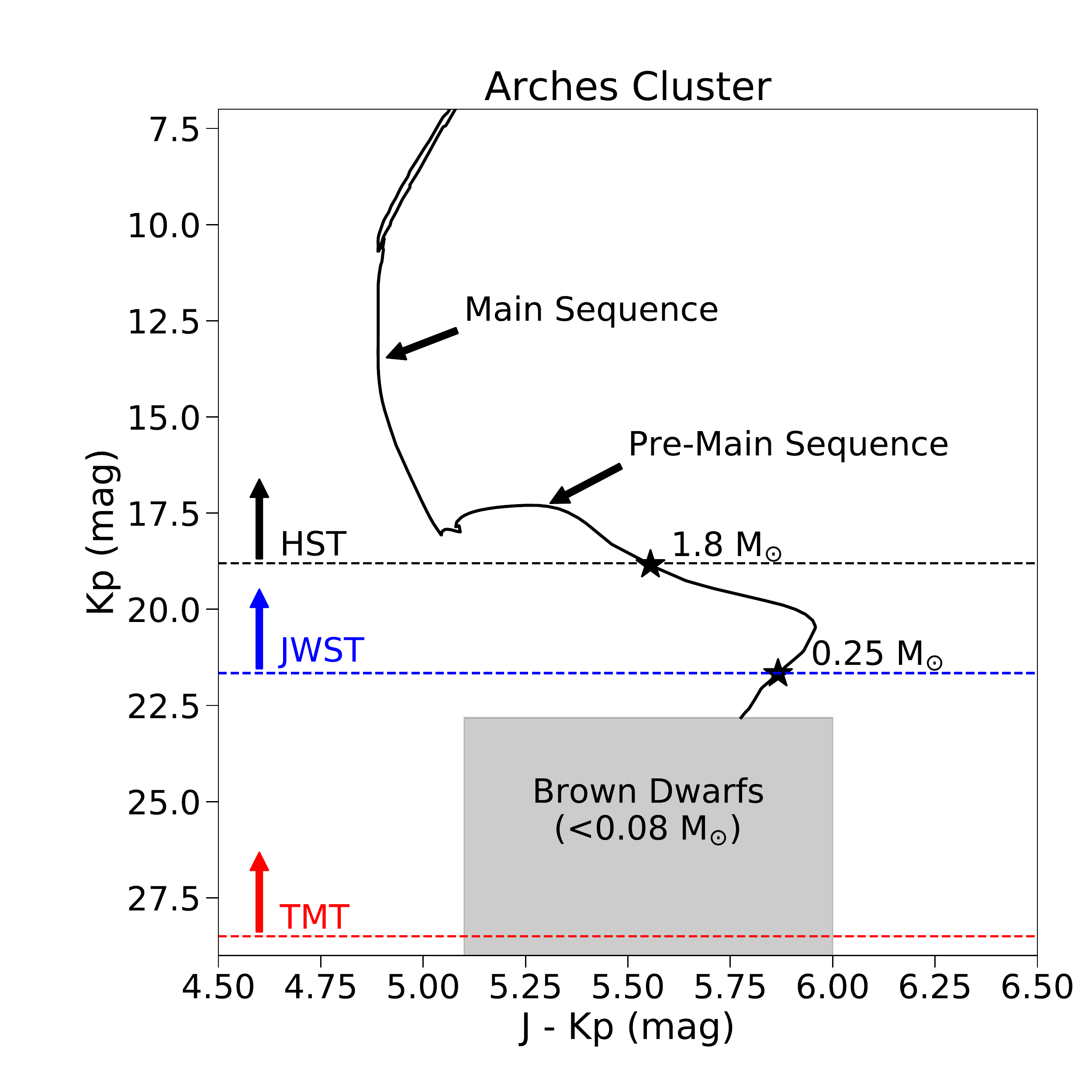}
        \caption{The TMT confusion limit (red dotted line) compared to the JWST confusion limit (blue dotted line) and HST completeness limit (black dotted line) for the Arches cluster, a $\sim$3 Myr old cluster near the Galactic Center. With 30m class telescopes, we can resolve objects well into the brown dwarf regime.}
         \label{fig:TMT_YNC}
    \end{minipage}
    \vspace{-0.5 cm}
\end{figure}

JWST, WFIRST, and 30m telescopes provide the ability to directly measure the IMF down to $\leq$0.2 M$_{\odot}$ across a range of environments in the Milky Way and Local Group. 
Of particular interest are young ($<$150 Myr) and massive ($>10^3$ M$_{\odot}$) clusters, which exhibit a single-age stellar population with a well-sampled IMF across a large mass range. 
Available targets include clusters in the disk and center of the Milky Way (e.g. Arches cluster, Westerlund 1), the LMC and SMC (e.g. R136, NGC 602), and M31 (e.g. KW258).
While the IMF has been measured to such depths in local star forming regions, these targets are significantly more massive and span a wide range of conditions including the Galactic Center, low metallicities, and starburst-like environments (Figure \ref{fig:sample}). 
These are highly crowded clusters and the lowest-mass stars are very faint, and so high spatial resolution and sensitivity are essential in order to make this measurement (Table \ref{tab:cluster_props}). 
For example, current resolved imaging studies of clusters at the Galactic Center are limited to the high-mass stars (M $>$ 2 M$_{\odot}$), at best.  
With 30m telescopes, the confusion limit is not reached until beyond the hydrogen-burning limit (Figure \ref{fig:TMT_YNC})! 
Wide-field observations by JWST and WFIRST allows for efficient coverage of the lower-density outer regions of Milky Way clusters which 
subtend $\gtrsim$1$^\prime$ on the sky.
\newline



\begin{figure}
   \vspace{-1cm}
   \centering
    \begin{minipage}{0.45\textwidth}
    \includegraphics[scale=0.32]{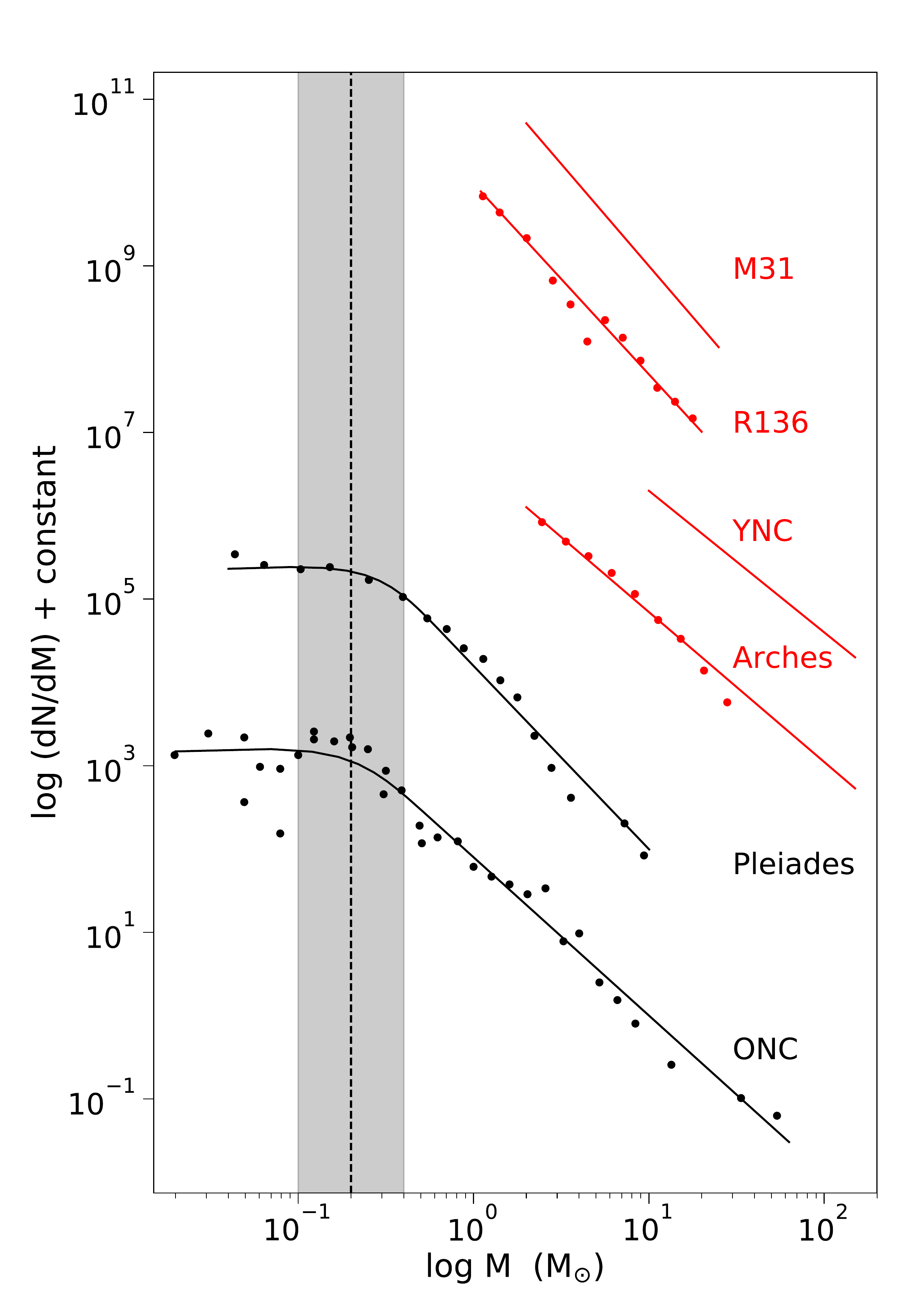}
    \end{minipage}\hfill
    \begin{minipage}{0.45\textwidth}
    \caption{Current IMF measurements for selected local star forming regions (black) and interesting future objects (red). With 30m class telescopes, we will be able to directly measure the IMF down to the peak mass (black shaded region) and beyond, characterizing its behavior across a wide range of environments. References: ONC \citep{Slesnick:2004fj, De-Marchi:2010yg}, Pleiades \citep{Moraux:2003ty, De-Marchi:2010yg}, Arches \citep{Hosek:2019kk}, YNC \citep{Lu:2013wo}, R136 \citep{Andersen:2009pl}, and M31 young clusters \citep{Weisz:2015lb}.}
    \label{fig:imf_plot}
    \end{minipage}
    \vspace{-0.5cm}
\end{figure}

Critically, achieving a depth of $\leq$0.2 M$_{\odot}$ allows us to accurately measure the peak of the IMF as well as the high-mass slope, as has been achieved for local star forming regions (Figure \ref{fig:imf_plot}).
For a typical 10$^4$~M$_{\odot}$ cluster, simulations by \citet{El-Badry:2017qf} show that this depth is required to distinguish between the proposed lognormal \citep[e.g.][]{Chabrier:2003wb} and broken power-law \citep[e.g.][]{Kroupa:2013qm} forms of the IMF, an uncertainty that hinders IMF measurements at low stellar masses today. 
In addition, assuming a log-normal IMF, they find that when observations reach 0.2 M$_{\odot}$ the degeneracy between the characteristic mass (i.e. peak mass) and width is broken and strong constraints can be placed on both parameters.
The peak of the IMF is a key constraint for star formation models since it is not a scale-free parameter (unlike the high-mass slope) and thus additional physics beyond gravity-driven accretion or turbulence is required to set it \citep{Krumholz:2014ne}. 
Possibilities include the thermal Jeans Mass \citep[e.g.][]{Larson:2005sp}, the turbulent Jeans Mass \citep[e.g.][]{Hennebelle:2008ht}, and radiative feedback \citep[e.g.][]{Bate:2009jw}. 
Predictions for how the IMF, and in particular the peak mass of the IMF, behaves in different environments changes depending on which of these processes dominate.
Thus, exploring these environments is a valuable tool for understanding star formation.  
\newline

\begin{deluxetable}{lccccc}
\tablewidth{0pt}
\tabletypesize{\footnotesize}
\tablecaption{Properties of Young Clusters in Different Environments \label{tab:cluster_props}}
\tablehead{
\colhead{Cluster} &
\colhead{Mass} &
\colhead{Distance} & 
\colhead{Angular} & 
\colhead{Proper Motion} & 
\colhead{0.2 M$_{\odot}$} \\
\colhead{} & 
\colhead{} &
\colhead{} &
\colhead{Diameter} &
\colhead{for 10 km s$^{-1}$} &
\colhead{Mag} \\
\colhead{} &
\colhead{(M$_{\odot}$)} &
\colhead{(kpc)} & 
\colhead{(arcsec)} & 
\colhead{(mas yr$^{-1}$)} & 
\colhead{(K mag)} 
} 
\startdata
Orion & $\sim$10$^3$ & 0.4 & 516 & 5.28 & 12.0 \\
Wd1 & 5x10$^4$ & 4.0 & 360 & 0.53 & 18.9 \\
Arches & 2.5x10$^4$ & 8.0 & 150 & 0.26 & 21.9 \\
R136 & 2.2x10$^4$ &   50 & 38.6 & 0.04 & 23.0  \\
KW258 & 1.1x10$^4$ & 753 & 2.45 & 3x10$^{-3}$ & 28.9 \\
\enddata
\vspace{-0.1in}
\end{deluxetable}

\begin{figure}
    \includegraphics[scale=0.35]{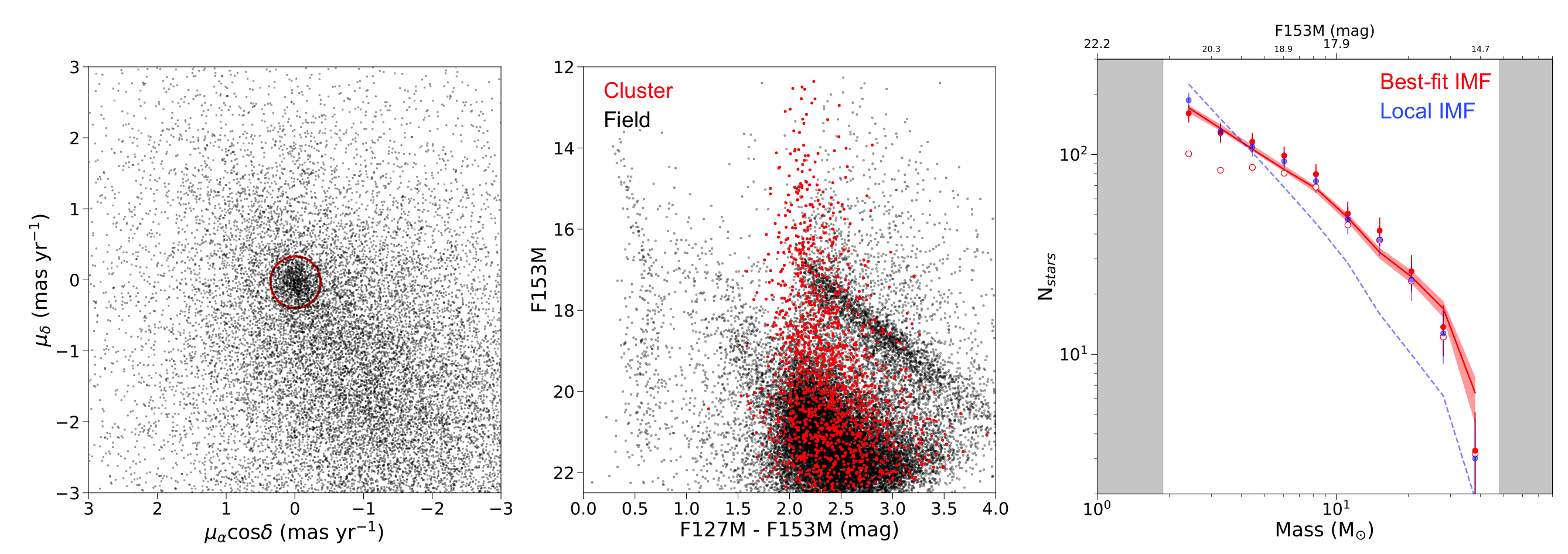}
    \caption{{\bf Left:} A proper motion diagram of the Arches cluster. Stars bound to the cluster have similar proper motions and thus form a well-defined clump (red circle), from which cluster membership can be determined. {\bf Middle:} The \emph{HST} color-magnitude diagram of the Arches cluster field, with cluster member candidates highlighted in red. The overlap between the cluster and field populations causes significant contamination without proper-motion selection. {\bf Right:} The Arches IMF (observations as red points, best-fit model as red shaded region) measured to down to 1.8 M$_{\odot}$ from \citet{Hosek:2019kk}. The cluster is found to have an overabundance of high-mass stars relative to nearby star forming regions (blue dotted line).}
    \label{fig:Arches_cmd}
\end{figure}

{\bf \large Methodology}
\newline
A common method to directly measure the IMF is to obtain the fluxes from the resolved stars, construct a color-magnitude diagram or luminosity function (correcting for differential reddening, if necessary), and then compare the population to theoretical isochrones from stellar evolution models. 
Due to their youth, the clusters are still partly embedded in the molecular cloud out of which they formed these studies are easiest performed in the NIR. 
Purely photometric studies of young clusters are often limited by field contamination, as differential extinction makes it difficult to identify cluster members via photometry alone. 
In addition, the fraction of field stars increases at fainter magnitudes, as the cluster luminosity function eventually turns over but the field star luminosity function continues to increase.
This issue can be mitigated by using proper motions in addition to photometry to isolate cluster members, as has been demonstrated for the Arches cluster \citep[Figure \ref{fig:Arches_cmd},][]{Hosek:2019kk}.  
\newline

IMF measurements can further be refined by measuring stellar temperatures from NIR spectra and constructing an HR diagram, which helps constrain the mass-luminosity relationship.
In addition, spectroscopy of a subset of cluster candidates can be used to quantify and correct for field contamination that remains after membership selection.
Spectral analysis can be performed using comparisons to both model spectra \citep[e.g.][]{Repolust:2005uq} and reference spectra from nearby star forming regions \citep[e.g.][]{Luhman:2016wk}.
These measurements at high stellar densities and large distance requires sensitive integral-field or multi-object spectrographs with R $\gtrsim$ 4000.
\newline


The IMF observations will produce cluster star catalogs containing photometry, proper motions, stellar masses, and membership probabilities for the resolved stars, as well as some spectroscopic measurements.
A large amount of ancillary science can be achieved with such data, including: 1) star cluster evolution and dynamics, taking advantage of the age spread across the clusters; 2) stellar evolution, as the high-quality cluster samples over a wide mass range at different ages and metallicities are a strong observational test for stellar evolution and atmosphere models; and 3) circumstellar disk evolution as a function of stellar mass, metallicity, and cluster environment. 
In addition, the clusters discussed here further serve as references themselves; objects such as 30 Doradus and its central cluster R136 are used as templates for the high-z unresolved star clusters where their properties have to be determined from their integrated properties. 
\newline
 
{\bf \large Recommendations}
\newline 
A {\bf thirty-meter diameter optical/NIR telescope equipped with adaptive optics} is the primary recommendation to advance observational studies of the stellar initial mass function over a wide range of environments. In particular:

\begin{itemize}
\item The adaptive optics system should provide wide field-of-view imaging capabilities with uniform correction in order to maximize effective observing area
\item The point-spread function will need to be known to high accuracy over the field, in order to take advantage of the huge gains provided by astrometry and proper motions
\item Moderate resolution (R $\sim$ 4000) spectroscopy from an integral field or multi-object infrared spectrograph, also supported by an adaptive optics system, is valuable to constrain the mass-luminosity relationship and quantify field contaminants
\item Having facilities in both hemispheres is required to take full advantage of available targets 
\end{itemize}

Such a facility has {\bf strong synergy with upcoming space telescopes such as JWST and WFIRST}. The large field of view provided by the space telescopes can efficiently observe the lower-density outer regions of a cluster, while the 30m telescopes can focus on the innermost crowded regions. Together, these observations ensure full spatial coverage of the target cluster.


\pagebreak
\bibliography{Thesis}

\end{document}